\documentclass[12pt]{article}

\usepackage[a4paper]{geometry}
\usepackage{amsthm}
\usepackage{amsmath}
\usepackage{amssymb}
\usepackage[dvips]{graphicx}
\usepackage{subfigure}

\newtheorem{lemma}{Lemma}
\newtheorem{theorem}{Theorem}

\begin{document}

\title{Fixed energy potentials through an auxiliary inverse eigenvalue problem}
\author{Tam\'as P\'almai and Barnab\'as Apagyi\\Department of Theoretical Physics, Institute of Physics\\
Budapest University of Technology and Economics\\ H-1111 Budafoki ut 8, Budapest, Hungary\\palmai@phy.bme.hu}
\maketitle
\begin{abstract}
An inverse scattering method based on an auxiliary inverse Sturm-Liouville problem
recently proposed by Horv\'ath and Apagyi [Mod. Phys. Lett. B 22, 2137 (2008)]
is examined in various aspects and developed further to (re)construct spherically
symmetric fixed energy potentials of compact support realized in the three-dimensional
Schr\"odinger equation. The method is generalized to obtain a family of inverse procedures
characterized by two parameters originating, respectively, from the Liouville
transformation and the solution of the inverse Sturm-Liouville problem. Both parameters
affect the bound states arising in the auxiliary inverse spectral problem and one of
them enables to reduce their number which is assessed by a simple method. Various solution
techniques of the underlying moment problem are proposed including exact Cauchy matrix inversion
method, usage of spurious bound state and assessment of the number of bound states.
Examples include (re)productions of potentials from phase shifts known theoretically or derived from scattering experiments.
\end{abstract}

\section{Introduction}

We consider the three-dimensional inverse scattering problem of the
Schr\"odinger equation on the half-line at a fixed scattering energy \cite{Newton, CS}.
For spherically symmetric potentials the partial wave expansion applies and the radial Schr\"odinger equations
\begin{equation}\label{radSch}
r^2\left[-\frac{d^2}{d r^2}+q(r)-k^2\right]\varphi_l(r)=-l(l+1)\varphi_l(r),\qquad l=0,1,2,\ldots
\end{equation}
with the physical boundary conditions describe the scattering of two non-relativistic
quantum mechanical objects (potential scattering). Presently $k$ is fixed to a constant value and we
assume that $q(r)$ is compactly supported, i.e. $q(r)=0$ for $r\geq a$. Moreover we require that
\begin{equation}\label{inL11}
rq(r)\in L^1(0,a).
\end{equation}

For this problem it can be shown that
\begin{eqnarray}
\label{phiin0}\varphi_l(r)=C_lr^{l+1}(1+o(1)),\qquad r\to0,\\
\varphi_l(r)=A_l\sqrt{r}\left(J_{l+1/2}(kr)-\tan\delta_l
Y_{l+1/2}(kr)\right),\qquad r\geq a,
\end{eqnarray}
where the phase shifts $\delta_l$ arise.

In \cite{HorvathTAMS} it was shown that a subset of the fixed energy phase
shifts, whose indices
satisfy the M\"unz condition,
determines the m-function and thus the spectral function, and thereby the potential of an
auxiliary inverse eigenvalue problem (i.e. an inverse Sturm-Liouville problem).
Based on this proof a constructive method was suggested in \cite{HAMPLB} for
the solution of the inverse scattering problem at fixed energy.

In this paper we develop this method further by generalizing both the
transformation of the fixed energy inverse scattering problem to the inverse
spectral problem and the constructive inversion method of the inverse eigenvalue
problem. In \cite{HAMPLB} only the case when one bound state is present in
the auxiliary problem has been addressed. We examine the zero and the multiple bound
state cases. In the latter one we find that it involves a highly nonlinear system of
equations which is difficult to solve. One way to overcome the difficulties
would be to reduce the number of bound states and to leave the nonlinear regime.
Remarkably, this can be achieved in many cases by tuning the parameters found by the
generalization of the original method. Also, an approximative argument is presented
to assess the number of bound states present in the auxiliary problem.

The paper is structured as follows: in the next section the transformation of
the fixed energy inverse scattering problem is generalized. Section 3 contains a
brief summary of the classical inverse Sturm-Liouville theory and the utilization thereof
in the present problem. In section 4 solution methods of the inverse problem
at different bound state levels are discussed. In section 5 an approximative
method is given to determine the number of bound states and also the possibility to
reduce their number is studied. Section 6 is devoted to illustrative applications,
section 7 is left for a summary.

\section{Liouville transformation}

First the fixed energy problem is transformed to an inverse eigenvalue problem
where the value of the m-function for some arguments is determined from the original fixed energy phase shifts.
To this end we transform the radial Schr\"odinger equation (\ref{radSch}) to the Liouville normal form
(see e.g. \cite{BRODE}) by using a Liouville transformation,
\begin{equation}
r\rightarrow x(r),\qquad \varphi_l(r)\rightarrow  \psi_l(x).
\end{equation}
Rewriting the differential equation (\ref{radSch}) in terms of the new independent variable $x$
and dependent variable $\psi_l(x)=f(x)^{-1}\varphi_l(r(x))$ yields
\begin{equation}
 -\psi_l''(x)-\left[\frac{\ddot x}{\dot
x^2}+2\frac{f'(x)}{f(x)}\right]\psi_l'(x)+\left[\frac{q(r(x))-k^2}{\dot x^2}-\frac{\ddot x}{\dot
x^2}\frac{f'(x)}{f(x)}-\frac{f''(x)}{f(x)}\right]\psi_l(x)=-\frac{l(l+1)}
{r(x)^2\dot x^2}\psi_l(x)
\end{equation}
where dot denotes differentiation with respect to $r$.
To get the Liouville normal form  of the Sturm-Liouville equation we need
\begin{equation}
r(x)^2\dot x^2={\rm const.}=c^2,
\end{equation}
and
\begin{equation}
\left[\frac{\ddot x}{\dot x^2}+2\frac{f'(x)}{f(x)}\right]\equiv0.
\end{equation}
The two conditions yield the unique solution
\begin{equation}\label{trafo}
x(r)=c\log r+c_1
\end{equation}
and
\begin{equation}
f(x)=c_2 e^{\frac{x}{2c}}.
\end{equation}
If we want to use the inverse spectral theory of the Sturm-Liouville equation
on $(0,\infty)$
we need $x(0)=+\infty$ and $x(a)=0$ which in turn implies
\begin{equation}
\rm{sgn}\, c=-1,\qquad c_1=-c\log a.
\end{equation}
Without the loss of generality $c_2=1$ is set and then the only remaining parameter $c$
can be chosen arbitrarily maintaining the negative sign.

In summary we have only one family of Liouville transformations reducing the radial
Schr\"odinger equation (\ref{radSch}) to the Liouville normal form, namely
\begin{eqnarray}
x(r)=c\log\frac{r}{a},\qquad c<0,\\
\psi_l(x)=e^{-\frac{x}{2c}}\varphi_l(ae^{\frac{x}{c}}).
\end{eqnarray}
Thereby (\ref{radSch}) transforms to
\begin{equation}\label{SLatlam}
-\psi_l''(x)+Q(x)\psi_l(x)=-\frac{1}{c^2}\left(l+\frac{1}{2}
\right)^2\psi_l(x),
\end{equation}
with the auxiliary potential
\begin{equation}\label{trpot}
Q(x)=\frac{a^2}{c^2}e^{\frac{2x}{c}}\left(q(ae^{\frac{x}{c}})-k^2\right).
\end{equation}
The transformed equation can be viewed as
\begin{equation}\label{StLio}
S[Q(x)]y(x,\lambda)=\lambda y(x,\lambda), \qquad S[Q(x)]=-\frac{d^2}{d
x^2}+Q(x)
\end{equation}
given explicitly at
\begin{equation}
\lambda=-\frac{1}{c^2}\left(l+\frac{1}{2}\right)^2
\end{equation}
also obtaining one of the two linearly independent solutions of the Sturm-Liouville equation as
\begin{equation}
y\left(x,-\frac{1}{c^2}\left(l+\frac{1}{2}\right)^2\right)=\psi_l(x).
\end{equation}
(Later we show that this is an $L^2$ solution.)

Note that in the original Horv\'ath-Apagyi \cite{HAMPLB} method $c=-1$ was taken implicitly.

\section{Inverse Sturm-Liouville problem}

\subsection{Summary of the classical inverse Sturm-Liouville problem}
\paragraph{Spectral properties}
For the Sturm-Liouville equation
\begin{equation}
-y_\alpha''(x,\lambda)+Q(x)y_\alpha(x,\lambda))=\lambda y_\alpha(x,\lambda),\qquad x\in [0,\infty)
\end{equation}
with the initial conditions
\begin{equation}
y_\alpha(0,\lambda)=\sin\alpha\neq0,\qquad y_\alpha'(0,\lambda)=-\cos\alpha,
\end{equation}
there exists \cite{Levitan,LS} a monotone increasing function
$\rho_\alpha(\lambda)$, the {\em spectral function}, such that, for every
$f(x)\in L^2(0,\infty)$ there exists in the $L^2(-\infty,\infty,\rho_\alpha(\lambda))$ norm sense
\begin{equation}
F_\alpha(\lambda)={\rm l.i.m.}_{n\to\infty}\int_0^n f(x)y_\alpha(x,\lambda)d x
\end{equation}
and this is a unitary transformation, i.e. the Parseval formula
\begin{equation}\label{Parseval}
\int_0^\infty |f(x)|^2d x=\int_{-\infty}^\infty
|F_\alpha(\lambda)|^2d \rho_\alpha(\lambda)
\end{equation}
holds (a theorem of Weyl).

A property of the spectral function is the formula \cite{Levitan,LS}
\begin{equation}\label{rhoexp}
\rho_\alpha(\lambda)=\frac{2}{\pi\sin^2\alpha}\lambda^{1/2}
+\rho_\alpha(-\infty)+\frac{\cos\alpha}{\sin^3\alpha}+o(1),\qquad
\lambda\to\infty.
\end{equation}

If $Q(x)\in L^1(0,\infty)$ then the Sturm-Liouville operator is in the
limit-point case at infinity and the
{\it{Weyl-Titchmarsh} m-function} is defined uniquely by
\begin{equation}
m(\lambda)=\frac{y'(0,\lambda)}{y(0,\lambda)}
\end{equation}
where $y(x,\lambda)$ is a solution belonging to the function space $L^2(0,\infty)$.

The m-function is related to the spectral function through a certain kind of Stieltjes transform \cite{Levitan}
\begin{equation}
\frac{\sin\alpha-m(\lambda)\cos\alpha}{\cos\alpha+m(\lambda)\sin\alpha}
=-\cot\alpha+\int_{-\infty}^\infty\frac{d\rho_\alpha(t)}{\lambda-t}.
\end{equation}

An equivalent formulation is given for the solution of the Sturm-Liouville equation with the initial conditions
\begin{equation}
y^h(0,\lambda)=1,\qquad y^{h}\,'(0,\lambda)=h<\infty.
\end{equation}
Prescribing $h=-\cot\alpha$ one has $[\sin\alpha\, y^h](0,\lambda)=\sin\alpha$ and $[\sin\alpha\, y^h]'(0,\lambda)=-\cos\alpha$
thus $\sin\alpha\, y^h(x,\lambda)=y_\alpha(x,\lambda)$. Then
$F_\alpha(\lambda)=\sin\alpha \int_0^\infty f(x)y^h(x,\lambda)d x=\sin\alpha F^h(\lambda)$ for $f(x)\in L^2(0,\infty)$ in the
norm sense. The Parseval formula (\ref{Parseval}) yields
\begin{equation}
\rho^h(\lambda)=\rho_\alpha(\lambda)\sin^2\alpha.
\end{equation}
$\rho^h(\lambda)$ is related to the m-function by
\begin{equation}
\frac{1}{m(\lambda)-h}=\int_{-\infty}^\infty\frac{d\rho^h(t)}{\lambda-t},
\end{equation}
which formula can be inverted by the Stieltjes inversion (see e.g. XIV. \textsection 3. of \cite{LS})
\begin{equation}\label{stieltjes}
\rho^h(\lambda_2)-\rho^h(\lambda_1)=-\frac{1}{\pi}
\lim_{\varepsilon\to0^+}\int_{\lambda_1}^{\lambda_2}{\rm
Im}\frac{1}{m(\lambda+i\varepsilon)-h}d\lambda.
\end{equation}

One can see that the two formulations are completely equivalent. We will use the latter one since it is traditionally used in the Gel'fand-Levitan construction discussed below.

We note that in the original formalism of Horv\'ath and Apagyi \cite{HAMPLB} $h=0$ was taken implicitly.

\paragraph{Construction of the potential from the spectral function}
From the existence of the spectral function the Gel'fand-Levitan (GL)
integral equation can be deduced \cite{Levitan}:
\begin{equation}\label{GL}
0=F(x,t)+K(x,t)+\int_0^x K(x,s)F(s,t)d s\qquad(0\le t\le x),
\end{equation}
where the input symmetrical kernel is
\begin{equation}
F(x,t)=\int_{-\infty}^\infty \cos(\sqrt{\lambda}x)
\cos(\sqrt{\lambda}t)d\sigma(\lambda)=\frac{1}{2}(F(x+t)+F(|x-t|)),
\end{equation}
\begin{equation}\label{Fdef}
F(x)=\int_{-\infty}^\infty
\cos(\sqrt{\lambda}x)d\sigma(\lambda),
\qquad\sigma(\lambda)=\rho^h(\lambda)-\rho^{0,0}(\lambda).
\end{equation}
$\rho^h(\lambda)$ is defined as before while $\rho^{0,0}(\lambda)$ is the spectral function
for the zero potential with boundary conditions $y(0)=1$, $y'(0)=0$, i.e.,
\begin{equation}
\rho^{0,0}(\lambda)=\begin{cases}\frac{2}{\pi}\sqrt{\lambda},& \lambda\geq0\\
0,& \lambda< 0.\end{cases}
\end{equation}

In the GL equation $K(x,y)$ is the kernel of the transformation operator
$T_{Q,0}:L^2\rightarrow L^2$ realized as
\begin{equation}
T_{Q,0}f(x)=f(x)+\int_0^xK(x,t)f(t)d t,\qquad
T_{Q,0}y_{0,0}(x,\lambda)=y_{Q,h}(x,\lambda)
\end{equation}
mapping the solutions of the Sturm-Liouville equation with $Q\equiv0$ satisfying
the boundary conditions
\begin{equation}
f(0)=1,\qquad f'(0)=0,
\end{equation}
onto the solutions with $Q\not\equiv0$ satisfying
\begin{equation}
f(0)=1,\qquad f'(0)=h<\infty.
\end{equation}
The transformation kernel is connected to the potential $Q(x)$ by
\begin{equation}
Q(x)=2\frac{d}{d x}K(x,x).
\end{equation}

\subsection{m-function of the operator $S[Q(x)]$ }

First we show that for the above-defined potential $Q(x)$, $Q(x)\in L^1(0,\infty)$
holds:
\begin{equation}
\int_0^\infty |Q(x)|d x\leq\frac{1}{c^2}\int_0^a
r^2|q(r)|d r+\frac{a^2k^2}{2|c|}<\infty
\end{equation}
by equation (\ref{inL11}), the fact that $\int_0^a
r^2|q(r)|d r<a\int_0^ar|q(r)|d r$ and $c<0$ was employed. Then $S[Q(x)]$ with
$Q(x)$ being the auxiliary potential is in the limit-point case.

Next we prove that the functions $\{\psi_l(x)\}_{l=0,1,\ldots}$ are of
the class $L^2(0,\infty)$:
\begin{equation}
\int_0^\infty |\psi_l(x)|^2d x=\int_0^\infty e^{-\frac{x}{c}} |\varphi_l(a
e^{\frac{x}{c}})|^2d x=a|c|\int_0^a \frac{|\varphi_l(r)|^2}{r^2} d r<\infty,
\end{equation}
where $l\geq0$ was exploited and the asymptotic formula
$\varphi(r)=Cr^{l+1}(1+o(1))$, $r\to0$ coming from equation (\ref{phiin0}) was used to
estimate the integral.

Since $\psi_l(x)\in L^2(0,\infty)$ and $Q(x)\in L^1(0,\infty)$ we infer that the m-function of the Sturm-Liouville operator $S[Q(x)]$ satisfies
\begin{equation}\label{mfunct}
m\left(-\frac{(l+1/2)^2}{c^2}\right)=\frac{\psi_l'(0)}{\psi_l(0)}=
\frac{ka}{c}\frac{J'_{l+1/2}(ka)-\tan\delta_l
Y'_{l+1/2}(ka)}{J_{l+1/2}(ka)-\tan\delta_l Y_{l+1/2}(ka)}.
\end{equation}
From this we can build the potential through the formula relating the m-function to the spectral function
and the constructive method discussed previously.

\subsection{Deriving a moment problem}

With reference to the defining formula (\ref{Fdef}) for $F(x)$ we define a truncated version
$\tilde{F}(x)$:
\begin{equation}
\tilde{F}(x)=\int_0^\infty \cos(\sqrt{\lambda}x)d\sigma(\lambda).
\end{equation}
It turns out that the reconstruction of the $\tilde F(x)$ function from the
given m-function values is an inverse moment problem. Consider
\begin{equation}
 I=\int_0^\infty\tilde{F}(x)e^{\left(l+\frac{1}{2}\right)\frac{x}{c}}d x=
\int_0^\infty\int_0^\infty
d\sigma(\lambda)d x\cos(\sqrt{\lambda}x)e^{\left(l+\frac{1}{2}\right)\frac{x}{c}}
=-\frac{1}{c}\int_0^\infty
d\sigma(\lambda)\frac{l+\frac{1}{2}}{\frac{1}{c^2}\left(l+\frac{1}{2}
\right)^2+\lambda}.
\end{equation}
This integration can be performed with ease in terms of the m-function and by
considering that on $(-\infty,0)$ $d\rho^h(\lambda)$ is concentrated at the bound states
$\lambda_1,\ldots,\lambda_B$ supported by $S[Q(x)]$:
\begin{align}
 I&=\frac{1}{c}\left(l+\frac{1}{2}\right)
\left[\int_{-\infty}^\infty
\frac{d(\rho^h(\lambda)-\rho^{0,0}(\lambda))}{-\frac{1}{c^2}\left(l+\frac{1}{2}
\right)^2-\lambda}
+\int_{-\infty}^0\frac{d\rho^h(\lambda)}{\frac{1}{c^2}\left(l+\frac{1}{2}
\right)^2+\lambda}\right]\\
&=\frac{1}{c}\left(l+\frac{1}{2}\right)
\left[\frac{1}{m\left(-\frac{1}{c^2}\left(l+\frac{1}{2}\right)^2\right)-h}-\frac
{1}{m_0\left(-\frac{1}{c^2}\left(l+\frac{1}{2}\right)^2\right)}+\sum_{i=1}^B\frac{
b_i}{\frac{1}{c^2}\left(l+\frac{1}{2}\right)^2+\lambda_i}\right].\,\,
\label{e45}
\end{align}
Here
\begin{equation}
b_i=\rho^h(\lambda_i+0)-\rho^h(\lambda_i-0)
\end{equation}
and $m_0(\cdot)$ denotes the m-function associated with $Q(x)\equiv0$.
The values of $m(\cdot)$ appearing in (\ref{e45}) are determined by
(\ref{mfunct}). For a more detailed derivation consult \cite{HAMPLB}.

Now we have the following problem for the truncated $\tilde F(x)$:
\begin{equation}
\int_0^\infty\tilde{F}(x)e^{\left(l+\frac{1}{2}\right)\frac{x}{c}}d x=\mu(c,h,
\delta_l,\{\lambda_i\},\{b_i\}),\qquad l=0,1,\ldots
\end{equation}
with the moments (assuming $B$ bound states)
\begin{multline}
\mu(c,h,\delta_l,\{\lambda_i\},\{b_i\})=\left(l+\frac{1}{2}
\right)\,\left(ka\,\frac{J'_{l+1/2}(ka)-\tan\delta_l
Y'_{l+1/2}(ka)}{J_{l+1/2}(ka)-\tan\delta_l Y_{l+1/2}(ka)}-c\,
h\right)^{-1}-1\\+\sum_{i=1}^B\frac{\frac{b_i}{c}\left(l+\frac{1}{2}\right)
}{\frac{1}{c^2}\left(l+\frac{1}{2}\right)^2+\lambda_i},
\end{multline}
which is essentially a moment problem. Note that unless there are no bound
states the moments depend on the undetermined quantities $\{\lambda_i\}$ and
$\{b_i\}$ associated to the bound state positions and norms.

\section{Solution method}
Depending on the number of bound states present in the auxiliary problem,
different solution strategies are called for. As a consequence it is important
to know or assess the number of bound states before solving the inverse problem which is
discussed in the next section.

\subsection{No bound states}

In this case we have a proper moment problem for $\tilde F(x)=F(x)$:
\begin{equation}\label{0BS}
\int_0^\infty
F(x)e^{\left(l+\frac{1}{2}\right)\frac{x}{c}}d x=\mu(c,h,\delta_l)\equiv\mu_l,
\end{equation}
where the moments now do not depend on unknown quantities.

To solve this moment problem we use the following expansion for $F(x)$:
\begin{equation}\label{powerseries}
F(x)=\sum_{n=0}^N c_n e^{-nx}.
\end{equation}
Upon substitution  into equation (\ref{0BS}) and using $N+1$ fixed
energy phase shifts as input data we get the following system of linear
equations for the coefficients:
\begin{equation}
\sum_{n=0}^N c_n \frac{-c}{-cn+l+\frac{1}{2}}=\mu_l,\qquad l=0,1,\ldots,N.
\end{equation}

Solving this system of linear equations yield the coefficients required to build
$F(x)$ and from that one can calculate the fixed energy potential essentially by
solving the GL integral equation (\ref{GL}). Note that solving the system of equations is not a
well-conditioned task as we must deal with a Hilbert-type matrix which is
infamously badly conditioned. Therefore, from the numerical point of view it is
of considerable
value to see, that this matrix can be inverted explicitly. Our matrix, i.e.
$\left[\frac{-c}{-cn+l+\frac{1}{2}}\right]_{ln}$ is in fact a Cauchy matrix. The
elements of the inverse of a general Cauchy matrix with elements
$a_{ij}=(x_i+y_j)^{-1}$ are given by \cite{CauchyMX}
\begin{equation}
b_{ij}=(x_j+y_i)\prod_{m\neq i} \frac{x_j+y_m}{y_m-y_i}\prod_{m\neq
j}\frac{x_m+y_i}{x_m-x_j}.
\end{equation}
In our case this implies
\begin{equation}
c_n=-\frac{1}{c}\sum_{l=0}^{N}\mu_l\,\left(l-cn+\frac{1}{2}\right)\prod_{
n'\neq n}\frac{l-cn'+\frac{1}{2}}{cn-cn'}\prod_{l'\neq l}
\frac{l'-cn+\frac{1}{2}}{l'-l},\qquad n=0,1,\ldots,N.
\end{equation}

\paragraph{Improvements of the solution method}
To further improve the solution method outlined above one can exploit two
properties of the $F(x)$ function, namely that
\begin{eqnarray}
F(0)=-h,\\
\lim_{x\to\infty}F(x)=0.
\end{eqnarray}
The latter is a simple consequence of the Riemann-Lebesgue lemma, however the
former needs more explanation. Let us write up
\begin{align}
 F(0)=\int_{-\infty}^\infty
d\sigma(\lambda)&=\lim_{\Lambda\to\infty}\int_{-\infty}^\Lambda
d\sigma(\lambda)=\lim_{\Lambda\to\infty}\left[
\rho^h(\Lambda)-\rho^h(-\infty)-\frac{2}{\pi}\sqrt{\Lambda}\right]=\nonumber\\&=
\lim_{\Lambda\to\infty}\left[\frac{2}{\pi}\sqrt{\Lambda}
+\rho^h(-\infty)-h+o(1)-\rho^h(-\infty)-\frac{2}{\pi}\sqrt{\Lambda}\right]=-h
\end{align}
where equation (\ref{rhoexp}) was used and continuity in $\Lambda$ was supposed.
This proves $F(0)=-h$.

Because of $F(\infty)=0$ one can take
\begin{equation}
c_0=0;
\end{equation}
however, we note that this is not always the best course of action in practical
scenarios (see Examples).

To incorporate the information $F(0)=-h$ there are several ways to choose from. For
instance, one can prescribe the condition
\begin{equation}
\sum_{n=0}^N c_n=-h
\end{equation}
for the coefficients (which complicates the solution process: the matrix to be
inverted is no longer of the Cauchy type). On the other hand, this will be very
useful in the one bound state case (see later) where it will permit to determine the
nonlinear parameter $\lambda$ in a linear way.

\paragraph{Hausdorff moment problem}
Our problem can be viewed as a Hausdorff moment problem since
with $z=e^{\frac{x}{c}}$ and $\mathfrak{F}(z)=-cz^{-1/2}F(c\log z)$ (\ref{0BS}) takes the form
\begin{equation}
\int_0^1 z^l\mathfrak{F}(z)d z=\mu_l,\qquad l=0,1,2,\ldots.
\end{equation}
The Hausdorff moment problem is studied in the literature in detail concerning both mathematical properties and solution
methods. For instance, an interesting stability result can be found in \cite{Talenti} whose corollary is the following theorem. It establishes an  accuracy estimate of the inverse moment problem which is  procedure independent.

\begin{theorem}\label{thm1}
Suppose the smoothness condition
\begin{equation}
\int_0^1 |\mathfrak{F}\,'(z)-\mathfrak{F}\,'_N(z)|^2dz\leq E^2 <\infty.
\end{equation}
Then if the first $N+1$ moments of $\mathfrak{F}(z)$ and $\mathfrak{F}_N(z)$ conincide, i.e.
\begin{equation}
\int_0^1 z^k \mathfrak{F}(z)dz=\int_0^1 z^k\mathfrak{F}_N(z)dz,\qquad k=0,1,\ldots N,
\end{equation}
we have
\begin{equation}
\int_0^1|\mathfrak{F}(z)-\mathfrak{F}_N(z)|^2dz\leq \frac{E^2}{4(N+1)^2}.
\end{equation}
\end{theorem}

Using Theorem \ref{thm1} one can conclude that if the moments $\mu_l$ are free of error, the difference between the approximated and the true $F(x)$ functions in $L^2$ norm tends to $0$ as the number of moments is increased:
\begin{equation}
\int_0^\infty |F(x)-F_N(x)|^2dx\leq\frac{C}{4(N+1)^2},
\end{equation}
with some $C$ constant depending on the smoothness of $F(x)-F_N(x)$,
\begin{equation}
\int_0^\infty |F'(x)-F_N'(x)|^2 e^{-\frac{2x}{c}}dx\leq C,
\end{equation}
and $F_N(x)$ is the approximation of the true $F(x)$ using the first $N+1$ moments.

It is interesting to consider the case when the input data is noisy. Using Theorem 1 of \cite{Talenti} one can state the following:
\begin{equation}
\int_0^\infty |F(x)-F_N(x)|^2dx\leq\min_n\left\{ \frac{\varepsilon^2}{|c|} e^{3.5(n+1)}+\frac{C}{4(n+1)^2}\, :\, n=0,1,\ldots N \right\},
\end{equation}
where $\varepsilon^2$ is the absolute square sum of the differences between the true and noisy moments. This result implies in particular that even if the number of phase shifts grows to infinity the recovery will not be complete when the data remains erroneous.

\subsection{One bound state}

Supposing (\ref{powerseries}) we get the system of equations
\begin{equation}
\sum_{n=0}^N c_n
\frac{-c}{-cn+l+\frac{1}{2}}=\mu_l+\frac{b\left(l+\frac{1}{2}\right)\frac{1}{c}}
{\frac{1}{c^2}\left(l+\frac{1}{2}\right)^2+\lambda},\qquad l=0,1,\ldots,N+1,
\end{equation}
where $\mu_l$ denotes the $l$th moment without the bound state contributions and $\lambda<0$ and $b>0$ are the bound state parameters (the subscript $1$ is omitted). Using
the expansion (\ref{powerseries}) for $\tilde F(x)$ we have
\begin{equation}
F(x)=b\cosh(\sqrt{-\lambda}x)+\sum_{n=0}^N c_n e^{-nx}.
\end{equation}
To get an explicitly solvable system of equations treating $\sqrt{-\lambda}$ as
a parameter we subtract the term $\frac{b}{2}e^{-\sqrt{-\lambda}x}$ from the expansion for $\tilde F(x)$, that is
\begin{equation}
\tilde F(x)=\sum_{n=0}^Nc_ne^{-nx}-\frac{b}{2}e^{-\sqrt{-\lambda}x},\qquad
F(x)=\frac{b}{2}e^{\sqrt{-\lambda}x}+\sum_{n=0}^N c_n e^{-nx},
\end{equation}
and obtain
\begin{equation}\label{1bound}
\frac{-c\, b}{2(\left(l+\frac{1}{2}\right)+c\sqrt{-\lambda})}+\sum_{n=0}^N
c_n \frac{-c}{-cn+l+\frac{1}{2}}=\mu_l,\qquad l=0,1,\ldots,N+1
\end{equation}
through the elementary identity
\begin{equation}
\frac{\alpha}{\alpha^2-\beta^2}-\frac{1}{2}\frac{1}{(\alpha+\beta)}=\frac{1}{2}
\frac{1}{\alpha-\beta},\qquad \alpha,\beta\in\mathbb{C}.
\end{equation}
The explicit solution of the system of equations is given by
\begin{multline}
c_n=-\frac{1}{c}\sum_{l=0}^{N+1}\mu_l\left(l-cn+\frac{1}{2}+c\delta_{n,-1}\sqrt{
-\lambda}\right)\prod_{n'\neq
n}\frac{l-cn'+\frac{1}{2}+c\delta_{n',-1}\sqrt{-\lambda}}{-cn'+c\delta_{n',-1}
\sqrt{-\lambda}+cn-c\delta_{n,-1}\sqrt{-\lambda}}\\
\times\prod_{l'\neq
l}\frac{l'-cn+\frac{1}{2}+c\delta_{n,-1}\sqrt{-\lambda}}{l'-l},
\end{multline}
where $n=-1$ is also allowed, $c_{-1}\equiv \frac{b}{2}$ and $\delta_{a,b}$ is
the Kronecker-delta. To determine $\sqrt{-\lambda}$ we use a result
concerning Cauchy matrices, namely that the sum of the coefficients (including
$b/2$) is linear in $\sqrt{-\lambda}$ (see e.g. \cite{HAMPLB}, Lemma 5):
\begin{equation}
S=\sum_{n=-1}^N c_n=\alpha \sqrt{-\lambda}+\beta.
\end{equation}
Using this fact the linear solution procedure is performed as follows.
\begin{enumerate}
\item Calculate $S$ for two arbitrarily chosen $\lambda$ values ($\lambda_{01}$ and $\lambda_{02}$)
solving (\ref{1bound}). Denoting the two sums $S_1$ and $S_2$, using the relation
$S=F(0)=-h$ get $\sqrt{-\lambda}$ by
\begin{equation}
\sqrt{-\lambda}=\frac{(S_2+h)\sqrt{-\lambda_{01}}-(S_1+h)\sqrt{-\lambda_{02}}}{S_2-S_1}.
\end{equation}
\item Calculate the $c_n$ coefficients from (\ref{1bound}) with the appropriate
$\lambda$
parameter obtained in (i).
\item Calculate $q(r)$ through $F(x)$ and the GL integral equation.
\end{enumerate}


\subsection{Multiple bound states}

 For two or more bound states we propose a non-linear problem as follows.
 \begin{enumerate}
 \item Solve the nonlinear system of equations
\begin{align}\label{2BS1}
&\sum_{i=1}^B\frac{-c\, b_i}{2(l+\frac{1}{2}+c\sqrt{-\lambda_i})}+
\sum_{n=0}^N c_n \frac{-c}{-cn+l+\frac{1}{2}}=\mu_l,\qquad l\in L\\
&\sum_{i=1}^B b_i+\sum_{i=0}^Nc_n=-h\label{2BS2}
\end{align}
with the index set $L=[0,1,\ldots,N+2B-1]$ in the variables
$\lambda_1,\lambda_2,\ldots,\lambda_B$, $b_1,b_2,\ldots,b_B$, $c_0,c_1,\ldots,c_N$.
\item Calculate $F(x)$ by
\begin{equation}F(x)=\sum_{i=1}^B\frac{b_i}{2}e^{\sqrt{-{\lambda_i}}x}+\sum_{n=0
}^N c_n e^{-nx}.\end{equation}
\item Solve the GL integral equation to obtain the potential.
\end{enumerate}

Numerically, it is worthwhile to initialize the variables for the nonlinear solver with the ones obtained  supposing a vanishing potential $q(r)\equiv0$. This is because, roughly speaking, $q(r)$ is expected to influence only the tail of the $Q(x)$ potential and the bound states shall be close to those corresponding to $q(r)\equiv0$ (see next section).

\section{Assessment of bound states}

We shall make observations based on the transformation formula for the potential (\ref{trpot}):
\begin{equation}
Q(x)=\frac{a^2}{c^2e^{\frac{2x}{|c|}}}\left[q(ae^{-\frac{x}{|c|}})-k^2\right],
\qquad q(a)=0.
\end{equation}

No bound state is expected when $q(r)>k^2$ on $0\leq r<a$, that is when $Q(x)$ is
positive everywhere, at least for $h=0$. This is the case, for example, when we have a
constant fixed-energy $q(r)$ potential in the form
\begin{equation}
q(r)=\begin{cases}C&\text{for } r < a,\\
0&\text{for } r\geq a,\end{cases}\qquad C>k^2.
\end{equation}

However, in practical applications it is a far more natural assumption that
$q(r)=0$ already on $b<r<a$ with some $b>0$. In this case we have
\begin{equation}
Q(x)=-\left(\frac{ka}{c}\right)^2e^{-2\frac{x}{|c|}},\qquad 0\leq
x<-|c|\log\left(\frac{b}{a}\right),
\end{equation}
and only the tail of $Q(x)$ is influenced by the fixed-energy potential
$q(r)$, e.g. $Q(0)$ is solely determined by $ka$. For this reason with given $k$ and
$a$ parameters by taking $q(r)\equiv0$ we can try to calculate the approximate
bound states or at least estimate their number. In what follows we
show exact results for the constant $q$-potential, which case contains the
zero $q$-potential case as well.

For $Q(x)=-s e^{-2t x}$ with $s,t >0$ the Sturm Liouville equation can be solved
explicitly. This is an easy exercise and will not be detailed here, only the
result is given.
The differential equation
\begin{equation}
-\psi''(x)-s e^{-2t x}\psi(x)=\lambda \psi(x)
\end{equation}
is solved by
\begin{equation}
\psi(x)=C_1 J_{i\sqrt{\lambda}/t}\left(\frac{\sqrt{s}}{t}e^{-tx}\right)+C_2
J_{-i\sqrt{\lambda}/t}\left(\frac{\sqrt{s}}{t}e^{-t x}\right)
\end{equation}
where Bessel functions generally of complex orders have appeared. For ${\rm
Im}\sqrt{\lambda}>0$ the $L^2$-solution (if $t>0$) is
\begin{equation}
\psi(x)=C_2 J_{-i\sqrt{\lambda}/t}\left(\frac{\sqrt{s}}{t}e^{-t
x}\right),
\end{equation}
since
\begin{equation}
J_{\pm i\sqrt{\lambda}/t}\left(\frac{\sqrt{s}}{t}e^{-t
x}\right)=b(x)x^{\mp{\rm Im}\sqrt{\lambda}/t}+o(x^{\mp {\rm
Im}\sqrt{\lambda}/t}),\qquad x\to0,
\end{equation}
where $b(x)$ is a bounded function. Then the m-function is
\begin{equation}
m(\lambda)=\frac{\psi'(0)}{\psi(0)}=-\sqrt{s}\frac{J_{-i\sqrt{\lambda}
/t}'(\sqrt{s}/t)}{J_{-i\sqrt{\lambda}/t}(\sqrt{s}/t)}.
\end{equation}
Using the Stieltjes inversion, equation (\ref{stieltjes}), $d\rho^h(\lambda)$ can
be found. For $\lambda>0$ trivially
\begin{equation}
d\rho^h(\lambda)=\frac{1}{\pi}{\rm
Im}\left[\sqrt{s}\frac{J'_{-i\sqrt{\lambda}/t}(\sqrt{s}/t)}{J_{-i\sqrt{
\lambda}/t}(\sqrt{s}/t)}+h\right]^{-1}d\lambda,\qquad \lambda>0.
\end{equation}

For $\lambda<0$ the measure is concentrated to points which are the bound states.
They are located at the eigenvalues of the operator where $\psi\in L^2(0,\infty)$.
Starting from $\psi(0)=1$ and $\psi'(0)=h$ one can calculate the coefficient of
the diverging solution to be
\begin{equation}
C_1=-\frac{\pi\sqrt{s}}{2t}\left[J'_{\sqrt{-\lambda}/t}
(\sqrt{s}/t)+\frac{h}{\sqrt{s}}J_{\sqrt{-\lambda}/t}
(\sqrt{s}/t)\right]
\end{equation}
where some elementary properties of the Bessel functions were used \cite{Watson}.
$C_1$ needs to be zero thus the bound states are located at $\lambda$'s that satisfy
\begin{equation}\label{BScond}
J'_{\sqrt{-\lambda}/t}(\sqrt{s}/t)+\frac{h}{\sqrt{s}}J_{\sqrt
{-\lambda}/t}
(\sqrt{s}/t)=0.
\end{equation}
Note that the number of bound states is greater than zero but finite.

Using the theory of residues the height of the step in $\rho(\lambda)$ at the bound
states can be obtained to be
\begin{equation}
\rho(\lambda_0+0)-\rho(\lambda_0-0)=2t\sqrt{-\lambda}\frac{J_{\sqrt{-\lambda_0}/t }(\sqrt{s}/t
)}{\sqrt{s}J^{(1,1)}_{\sqrt{-\lambda_0}/t }(\sqrt{s}/t )+hJ^{(1,0)}_{\sqrt{-\lambda_0}/t }(\sqrt{s}/t )},
\end{equation}
where the superscript $(n,m)$ means derivation with respect to order $n$ times and derivation with respect to the variable
$m$ times.

In case of $h=0$ we have a particularly simple scenario at hand. For definiteness let
\begin{equation}
\kappa^2=k^2-q(0)
\end{equation}
($q(0)$ being the value of the constant potential at the origin, and suppose for simplicity that $q(0)<k^2$). The potential $Q(x)=-\frac{(\kappa a)^2}{c^2} e^{-\frac{2x}{|c|}}$ with $h=0$ has bound states $\lambda_i$ at
\begin{equation}
J'_{|c|\sqrt{-\lambda_i}}(\kappa a)=0.
\label{assessment}\end{equation}
From \cite{Watson} we infer, denoting the $n$th root of $J'_\mu(x)$ by
$j'_{\mu,n}$, that $j'_{\mu,n}<j'_{\mu+\varepsilon,n}$ $n=1,2,\ldots$ holds with
$j'_{0,1}=0$. Moreover, $j'_{\mu,n}$ is continuous in $\mu$ \cite{Watson}; therefore,
we find that for $\kappa a>0$ there always exits at least one bound state. From
the above fact it also follows that the number of zeros of $J'_0(x)$ on $0\leq
x<\kappa a$ equals the number of bound states of
$Q(x)=-\frac{(\kappa a)^2}{c^2} e^{-\frac{2x}{|c|}}$. Alternatively, as
$J_0'(x)=-J_1(x)$, the number of bound states increase at the zeros of $J_1(x)$,
i.e. at $j_{1,n}$, $n=1,2,\ldots$. Some numerical results are listed in Table \ref{tab1}.
For a general potential $q(r)$ the data listed in Table \ref{tab1} can still be relevant
if the potential is shallow comparad to $k^2$ and only the tail of $Q(x)$ is influenced by $q(r)$.
Note that the value of the parameter $c$ does not affect the number of bound states
(only their positions) as it is only a scale parameter of the function
$J'_{|c|\sqrt{-\lambda}}(\kappa a)$ of $\sqrt{-\lambda}$ whose zeros give the bound states.

\begin{table}
\caption{\label{tab1}The first few sectors of definite bound state numbers for a constant $q(r)$ potential at $h=0$.}

\begin{center}
\begin{tabular}{cc}
\hline
$\phantom{3.831}0<\kappa a<3.8317$&one bound state\\
$3.8318<\kappa a<7.0156$&two bound states\\
$7.0156<\kappa a<10.174$&three bound states\\
$10.174<\kappa a<13.324$&four bound states\\
&etc.\\
\hline
\end{tabular}
\end{center}
\end{table}

Allowing $h\neq0$ gives rise to the more involved condition (\ref{BScond}) for the bound states.
It is possible then for some $h\neq0$ that one has one bound state while for $h=0$ two of them. This has the
favorable consequence of reducing a nonlinear problem to a linear one. Without giving an exhaustive treatment of the situation
we show the following illustrative result for bound state reduction.

\begin{lemma}
For $3.83\approx j'_{0,2}<\kappa a<j_{0,2}\approx 5.52$ the two bound states present at $h=0$ can be reduced to one
by varying $h$.
\end{lemma}

\begin{figure}
\includegraphics[width=10cm]{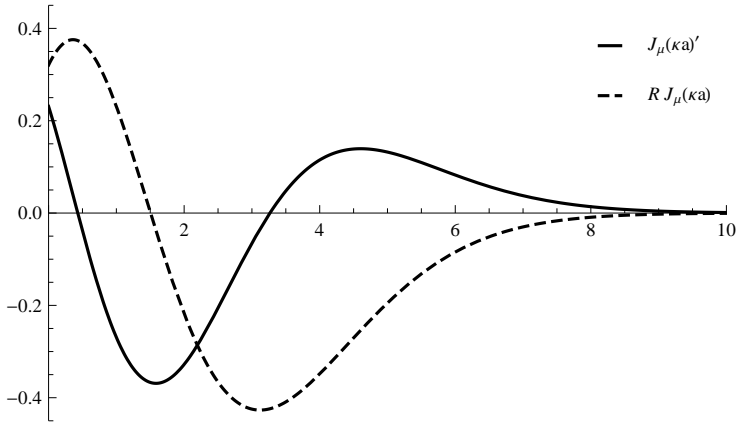}
\caption{\label{fig1} $J_{\mu}(\kappa a)'$ (full line) and $RJ_\mu(\kappa a)$ (dashed line) as functions of the order ($R$ is
a real number satisfying ${\rm sgn} (J_0'(\kappa a))={\rm sgn} (RJ_0(\kappa a))$ and $|RJ_0(\kappa a)|>|J_0'(\kappa a)|$).
While at $h=0$ the zeros of the derivative functions are the bound states, at $h\neq0$ the intersection of the two
graphs give them.}
\end{figure}

\begin{proof}
In figure \ref{fig1} the two Bessel-type functions entering condition (\ref{BScond}) are depicted.
We will show that by varying $h$ (on the figure through $R$), disregarding an overall sign, one can
always get the same kind of graphs as the ones on the figure.

By using the fact that the zeros of $J_0(x)$ and $J'_0(x)$ interlace we infer $J_{0}(\kappa a)\neq0$
and $J'_{0}(\kappa a)\neq0$. The only thing that remains to be shown is that the distribution of the zeros
are as depicted. Since $j'_{0,2}<\kappa a<j_{0,2}$ for $J'_k(\kappa a)$ only the first two and for $J_k(\kappa a)$ only
the first zeros can enter our considerations. Then the interlacing relation $j'_{a,1}<j_{a,1}<j'_{a,2}$
translates to $k_1<k_2<k_3$ if $j'_{k_1,1}=\kappa a$, $j_{k_2,1}=\kappa a$ and $j'_{k_3,2}=\kappa a$.
This completes the proof of the lemma.
\end{proof}

\section{Examples}

\subsection{Reconstruction of constant potentials}

First we reconstruct a potential
\begin{equation}
q(r)=\begin{cases}1.2&\text{for } r\leq 2= a\\
0&\text{for } r > 2=a\end{cases}
\end{equation}
at $k=1$ that generates no bound states in the auxiliary problem (because $Q(x)>0$).

\begin{figure}[ht]
  \begin{minipage}[b]{0.5\linewidth}
     \centering
     \subfigure[$5$ phases.]{\includegraphics[width=7cm]{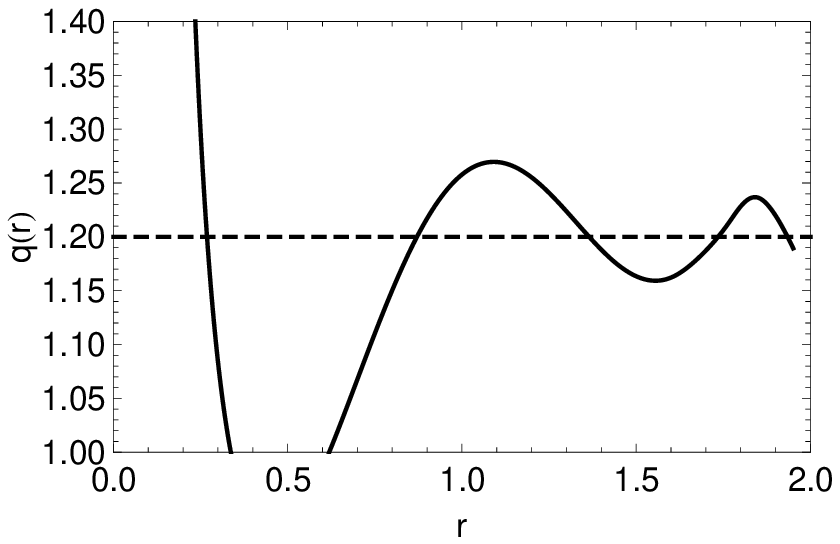}}\\
     \subfigure[$10$ phases.]{\includegraphics[width=7cm]{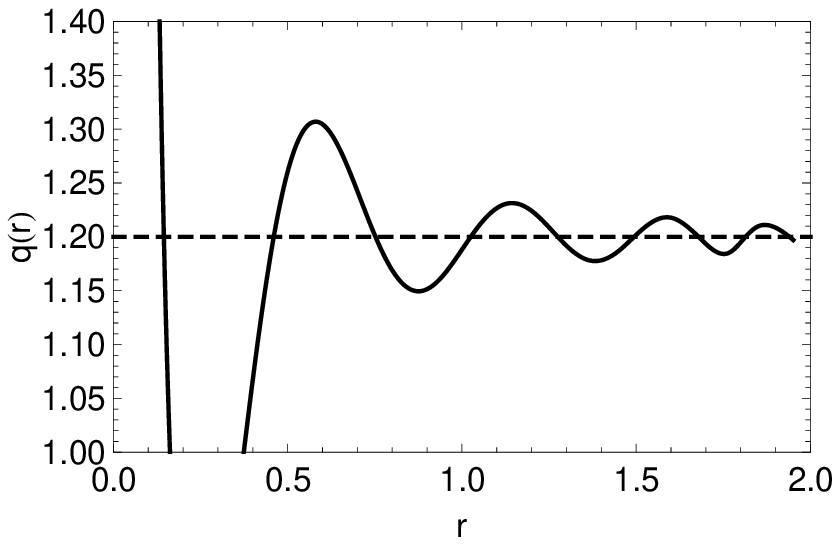}}\\
  \end{minipage}
  \hspace{0.1cm}
  \begin{minipage}[b]{0.5\linewidth}
     \centering
     \subfigure[$20$ phases.]{\includegraphics[width=7cm]{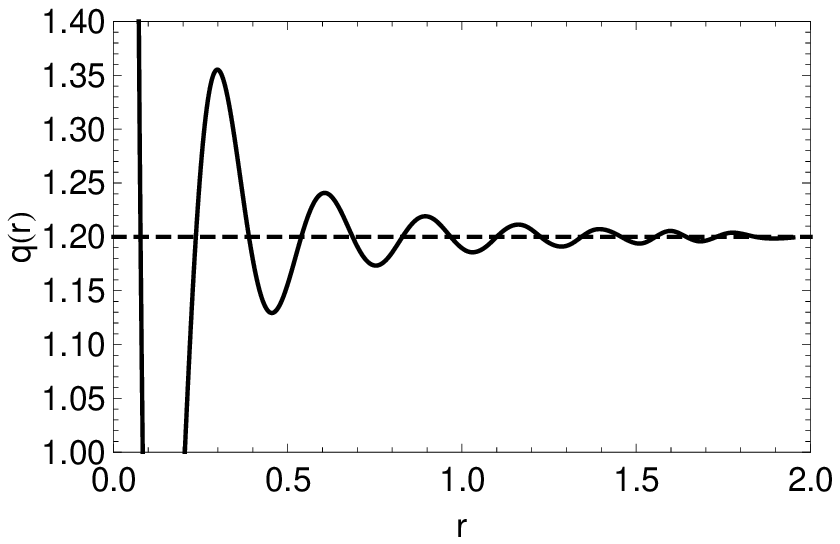}}\\
     \subfigure[$40$ phases.]{\includegraphics[width=7cm]{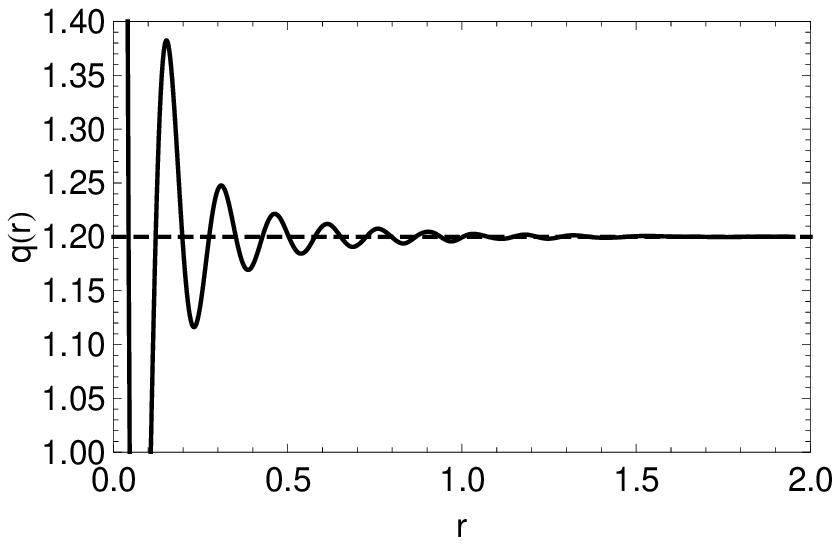}}\\
   \end{minipage}
   \caption{Reconstructions with different number of input phase shifts of the constant potential $q(r)=1.2\, H_2(r)$.
($c=-1$, $h=0$.)}
 \label{figN}
\end{figure}

Figure
\ref{figN}\footnote{For brevity of the figure captions we introduce the function $H_a(x)$
being a step function: $H_a(x)=1$ for $x\leq a$ and $H_a(x)=0$ for $x> a$.} shows the
quality of the reconstructions when $5$, $10$, $20$ and $40$ input phase shifts are
used with the standard choice of parameters ($c=-1$, $h=0$).
As expected we see that the quality of the inversion procedure gets better as the number of input data is growing.

\begin{figure}[ht]
  \begin{minipage}[b]{0.5\linewidth}
     \centering
     \subfigure[$c=-1$, $h=0$ (standard choice). s=12.]{\includegraphics[width=7cm]{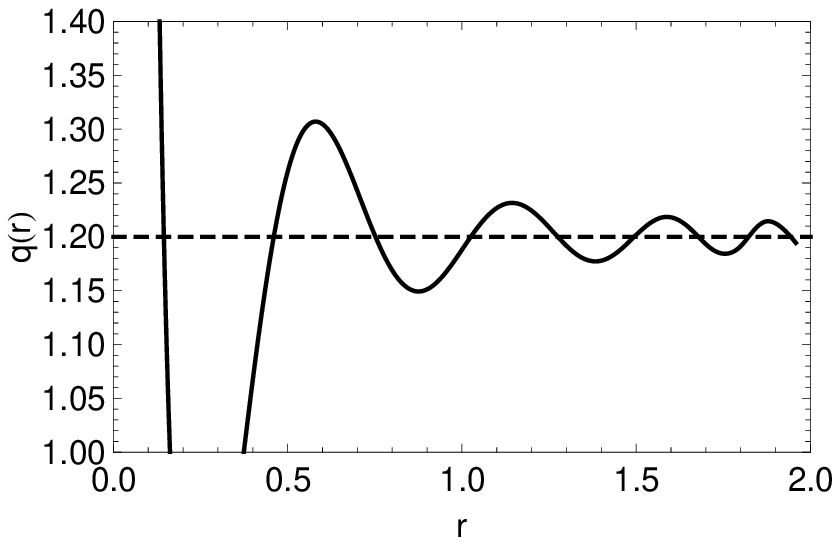}}\\
     \subfigure[$c=-1$, $h=-0.15$. s=15.]{\includegraphics[width=7cm]{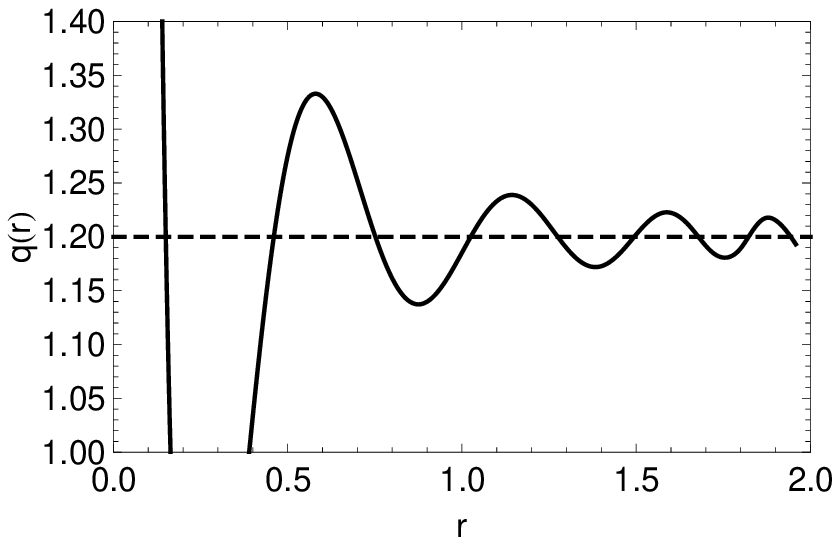}}\\
  \end{minipage}
  \hspace{0.1cm}
  \begin{minipage}[b]{0.5\linewidth}
     \centering
     \subfigure[$c=-0.30$, $h=0$. s=0.15.]{\includegraphics[width=7cm]{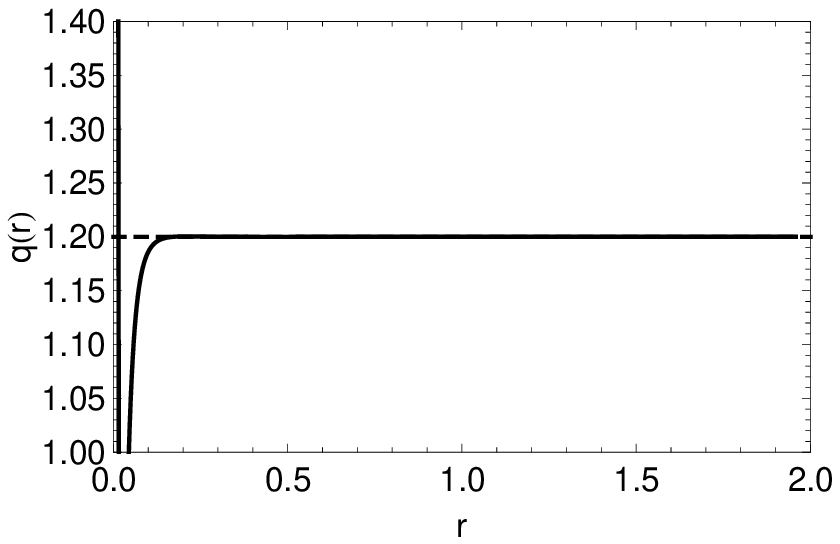}}\\
     \subfigure[$c=-0.30$, $h=-0.15$. s=0.011.]{\includegraphics[width=7cm]{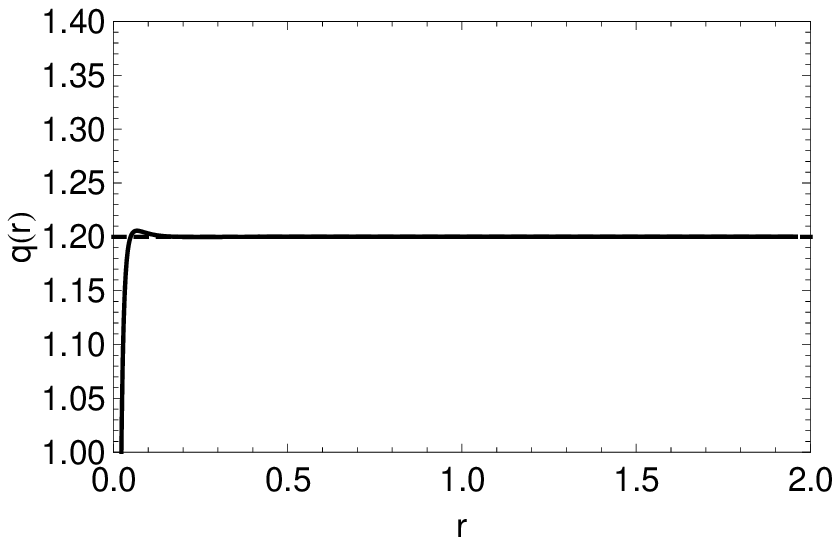}}\\
   \end{minipage}
   \caption{Reconstructions with different choices of the parameters of the constant potential $q(r)=1.2\, H_2(r)$
from 11 phase shifts with no bound states in the auxiliary problem.} \label{fig2}
\end{figure}

We may let the parameters $c$ and $h$ change from the standard values and prescribe the smoothness of the inverse potential $q(r)$. This can be done by monitoring the quantity
\begin{equation}
s(c,h)=\int_{r_0}^a |q'(r)|dr
\end{equation}
at each values of the parameters $c$ and $h$ where the lower limit $r_0$ is introduced in order to exclude (possible pathological)
singularity at the origin. Note that we used $r_0=0.05$ in all the examples shown.
By requiring  $s$ to be small is equivalent to exclude nonphysical potentials (see also \cite{PAJMP}). On the other hand, this requirement also means that we encounter smooth $F$-functions too, in agreement with Theorem \ref{thm1}.

Figure \ref{fig2} shows that choosing a modest number $N=10$ of input data we also get excellent reconstruction when $s$ is small. The optimal choice of parameters has proved to be $c=-0.3$ and $h=-0.15$ (\ref{fig2}(d)).

Several other techniques can also be employed to find optimal results. One is to retain the
constant term $c_0$ in the expansion (\ref{powerseries}). The value of $c_0$ (i.e. the departure from zero) serves also as an overall indicator of the accuracy of the procedure.
Another trick is to reconstruct the potential using the one bound state procedure, that is by retaining the $c_{-1}=b/2$ term and the associated nonlinear parameter $\sqrt{-\lambda}$ (a possible spurious bound state). This procedure also affects beneficently the numerics and can be applied both in the zero and one bound state cases.

\begin{table}
\caption{\label{tab2} Input phases $\delta_l$ and the intermediate quantities of the inversion procedure of the potential $q(r)=1.2\,H_2(r)$: moments $\mu_l$, coefficients $c_n$ and the nonlinear parameter (i.e. 'bound state').}

\begin{center}
\begin{tabular}{cccccc}
\hline
$l$&$\delta_l$&$\mu_l$&&$n$&$c_n$\\
\hline
$0$  &$-0.9890$&$-0.1714$&&$-1$&$-6.4667^{\rm a}$\\
$1$  &$-0.2964$&$-0.0043$&&$\phantom{-}0$&$+0.0002$\\
$2$  &$-0.0471$&$\phantom{-}0.0151$&&$\phantom{-}1$&$-0.0954$\\
$3$  &$-0.0037$&$\phantom{-}0.0180$&&$\phantom{-}2$&$+7.2085$\\
$4$  &$-0.0001$&$\phantom{-}0.0176$&&$\phantom{-}3$&$-0.1510$\\
$5$  &$-5.0\times 10^{-6}$&$\phantom{-}0.0164$&&$\phantom{-}4$&$+0.2657$\\
$6$  &$-1.1\times 10^{-7}$&$\phantom{-}0.0151$&&$\phantom{-}5$&$-0.2968$\\
$7$  &$-1.8\times 10^{-9}$&$\phantom{-}0.0139$&&$\phantom{-}6$&$+0.2027$\\
$8$  &$-2.2\times 10^{-11}$&$\phantom{-}0.0129$&&$\phantom{-}7$&$-0.2022$\\
$9$  &$-2.3\times 10^{-13}$&$\phantom{-}0.0119$&&$\phantom{-}8$&$+0.0258$\\
$10$&$-1.9\times 10^{-15}$&$\phantom{-}0.0111$&&$\phantom{-}9$&$+0.0092$\\
\hline
\end{tabular}\\
$^{\rm a}$ Nonlinear "bound state" parameter: $\sqrt{-\lambda}=-1.4447$.
\end{center}
\end{table}

In Table \ref{tab2} we list the input phase shifts $\delta_l$ and the intermediate data (moments $\mu_l$, coefficients $c_n$
 and the value  $\sqrt{-\lambda}$) of the inverse calculation with
  $c=-0.3$ and $h=-0.5$ (where $s=0.0004$). As we see the procedure yields
$\sqrt{-\lambda}=-1.4447$ for the "bound state" parameter, a negative value which clearly indicates that
there is no true bound state in the auxiliary problem.

\begin{figure}[ht]
\begin{center}
     \subfigure[$c=-1.0$, $h=0$, $\lambda=-0.105$, $s=0.049$.]{\includegraphics[width=7cm]{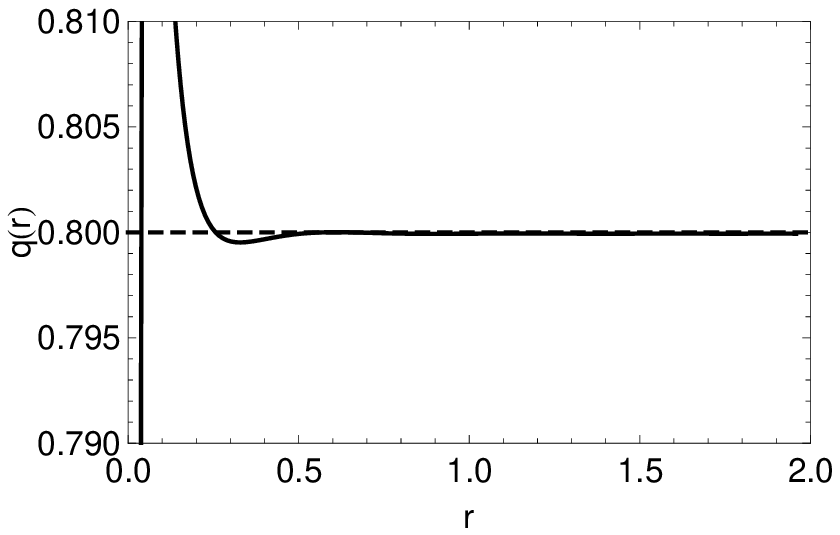}}$\qquad$
     \subfigure[$c=-0.5$, $h=-0.65$, $\lambda=-1.39$, $s=0.0022$]{\includegraphics[width=7cm]{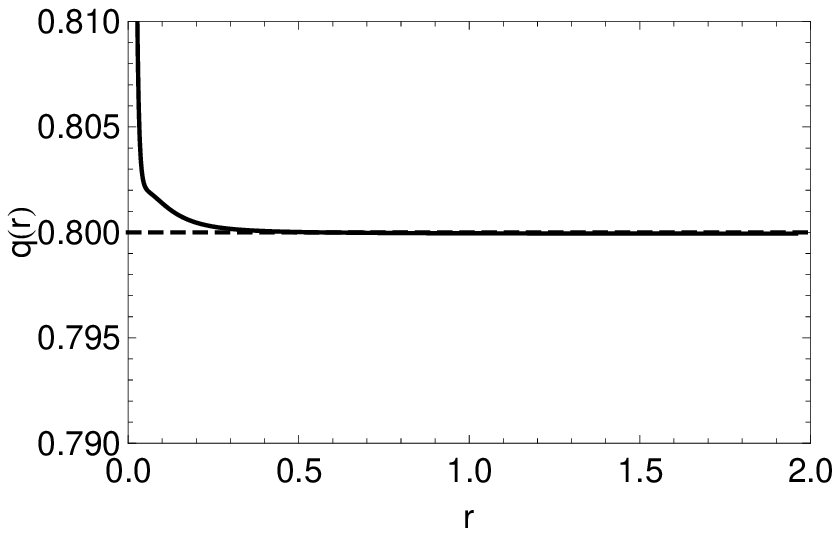}}\\
     \subfigure[$c=-1.5$, $h=0$, $\lambda_1=-0.48$, $\lambda_2=-2.43$, $s=1.28$, two bound states.]{\includegraphics[width=7cm]{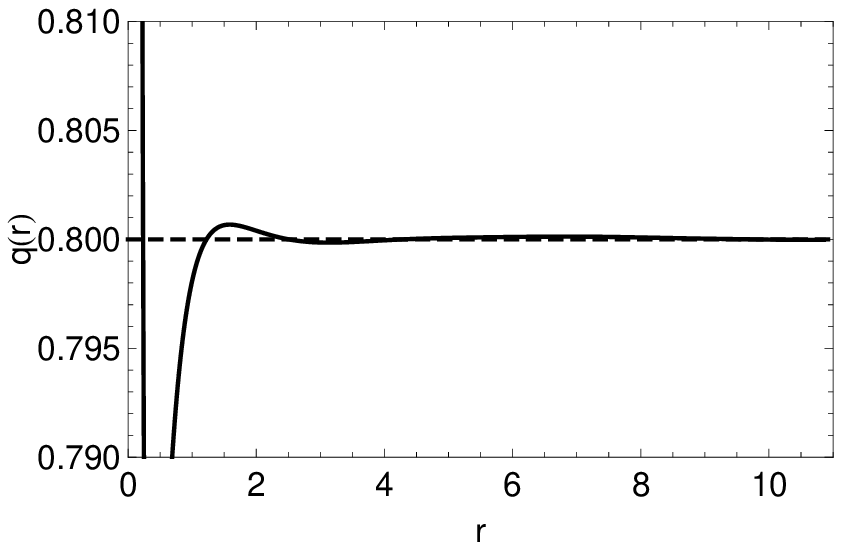}}$\qquad$
\subfigure[$c=-1.5$, $h=5$, $\lambda=-2.41$, $s=0.11$, one bound state.]{\includegraphics[width=7cm]{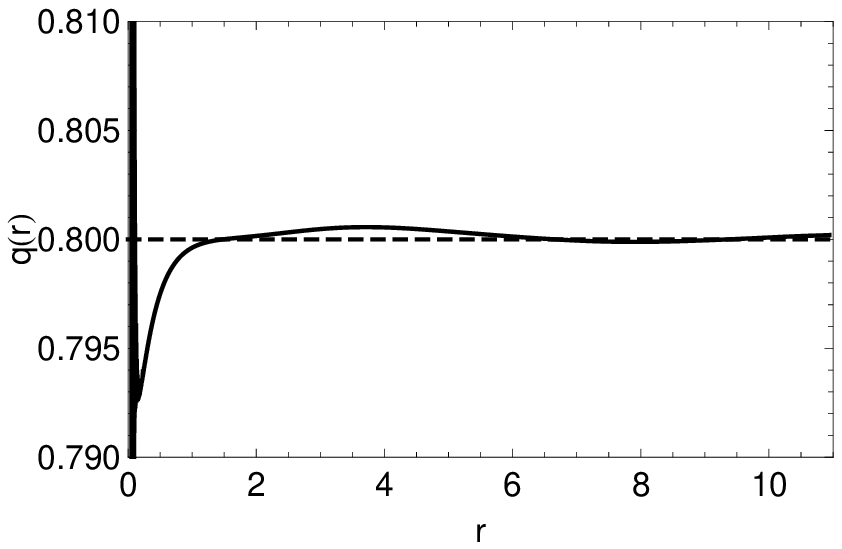}}
     \\
   \caption{(a)-(b) Reconstructions of the constant potential $q(r)=0.8\, H_2(r)$
from 11 phase shifts and one bound state in the auxiliary problem at $h=0$ with different choices of the parameters.
(c)-(d) Reconstructions of the potential $q(r)=0.8\, H_{11}(r)$ using (c) the two bound state formulation
($h=0$) and (d) the one bound state method ($h=5$). The constant term was present
in (\ref{powerseries}) during the calculation.
} \label{fig4}
\end{center}
\end{figure}

Now we proceed to reconstruct the potential
\begin{equation}
q(r)=\begin{cases}0.8&\text{for } r \leq  a\\
0&\text{for } r > a\end{cases}
\end{equation}
at $k=1$ that generates
 one auxiliary bound state with $
 a=2$ and two auxiliary bound states with $a=11$ (choosing $h=0$).

 Figure \ref{fig4}(a)-(b) shows that the inversion procedure yields a better potential for the smaller $s$ value
 also in this one (auxiliary) bound state case. The deviation of the inverse potential from the original one in the region $0.5< r < 2$ is of the
order of $0.00010$ for $s=0.049$ (at $c=-1$, $h=0$) and $0.00005$ for  $s=0.0022$ (at $c=-0.5$, $h=-0.65$).

Figure \ref{fig4}(c)-(d) shows two reconstructions in the case when two  bound states exist in the auxiliary problem.
In the first case (figure \ref{fig4}(c)) the inverse potential is obtained by the use of the two bound state method with parameters $c=-1.5$, $h=0$ and finding the nonlinear (bound state) parameters at $\lambda_1\approx - 0.48,\lambda_2\approx - 2.43$ (starting from the trial values of  $\lambda_1\approx - 0.8,\lambda_2\approx - 2.3$). In the second case
 (figure \ref{fig4}(d)), the inverse potential has been calculated by using the one bound state procedure with the parameters choice $c=-1$, $h=5$ (providing for the bound state parameter the value $\lambda=-2.41$).
We see that both methods yield similar results. However, while the two bound state calculation needs an a priori guess about the positions of the auxiliary bound states,  the one bound state method does not suffer from such an ambiguity and the calculation can be performed in the linear regime. One can check that in this case $\kappa a\approx 4.92$, thus we are in the domain where the number of bound states can be reduced (Lemma 1) by varying $h$.

\subsection{Reconstruction of potentials with different shapes}

\begin{figure}[ht]
\begin{center}
\subfigure[Gauss potential. $s=4.9$.]{\includegraphics[width=7cm]{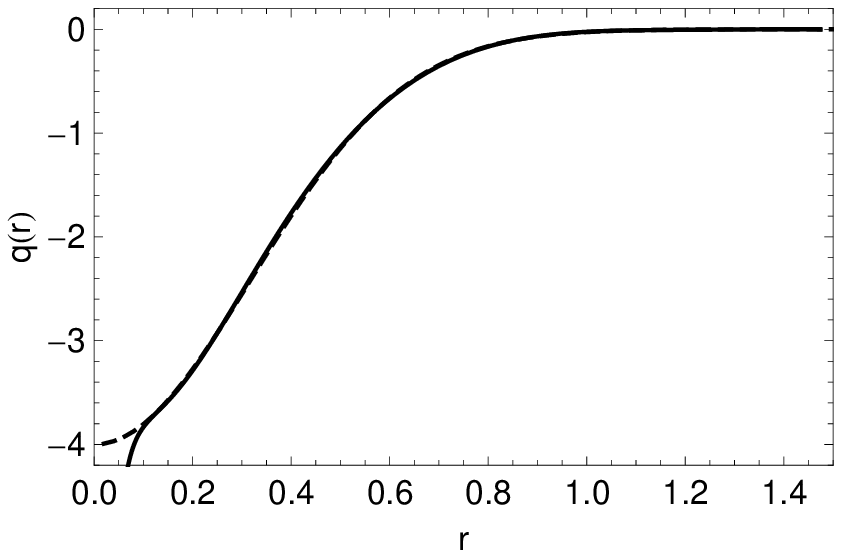}}$\qquad$
\subfigure[Woods-Saxon potential. $s=7.4$.]{\includegraphics[width=7cm]{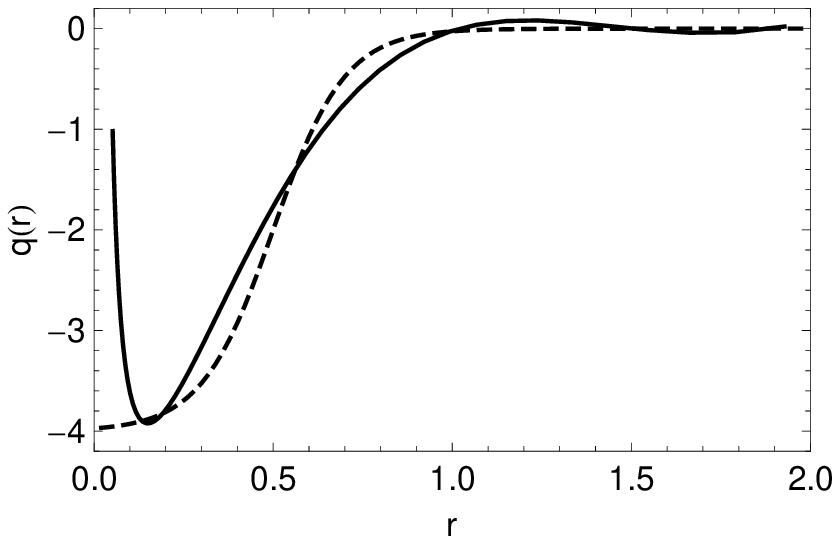}}
     \\
     \caption{Reconstruction of potentials with different shapes. (a) Gaussian shape from 7 phase shifts, (b) Woods-Saxon shape from 4 phase shifts.  The original potential is depicted with dashed line while the reconstruction is illustrated with solid line.  For  parameters and accuracy see the text.} \label{fig8}
\end{center}
\end{figure}

 Figure \ref{fig8}(a)-(b) shows
the reconstruction of the two potentials given by
\begin{eqnarray}
q_{\rm Gauss}(r)=-4e^{-5r^2},\qquad
q_{\rm WS}(r)=-\frac{4}{1+e^{\frac{r-0.5}{0.1}}}.
\end{eqnarray}
The reconstructions have been carried out at $k=1.5$ with 7 input phase shifts given with a precision of 4 digits  for the Gauss potential $q_{\rm Gauss}(r)$, and with 4 input phase shifts given with a  precision of only 2 digits for the Woods-Saxon shaped potential $q_{\rm WS}(r)$. In both cases the one bound state approximation procedure is applicable. For the Gauss potential the reconstruction with the parameters $a = 1.5, c = -0.74$, and $h = 0$ resulted in the bound state position at $\lambda=-3.36$. Note that this figure agrees with the exact bound state position (when calculated from the known auxiliary potential $Q(x)$) and only slightly differs form the assessment value of $\lambda=-3.22$ (calculated from equation (\ref{assessment})). For the WS potential the reconstruction was carried out with the parameters $a = 2, c = -1.25, h = 0 $ and resulted in the bound state position at $\lambda=-2.46$ to be compared with the assessed value of $\lambda=-2.44$.

\subsection{Inverse potentials from experimental phase shifts}
\subsubsection{ $e-Ar$ atom scattering at $E=12$ eV ($0.4412$ au).}

\begin{figure}[ht]
\begin{center}
\subfigure[$e-Ar$ potentials.]{\includegraphics[width=7.5cm]{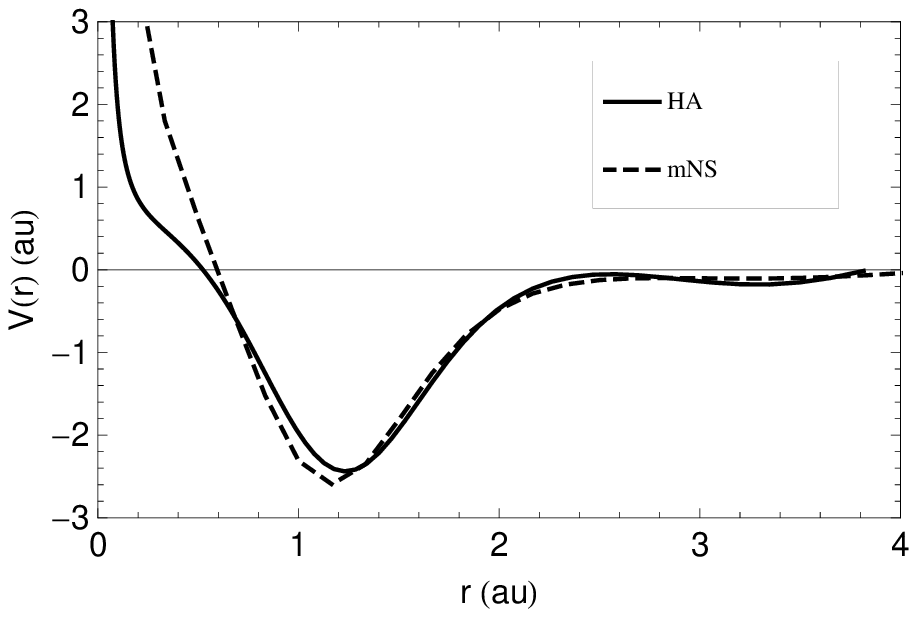}}$\qquad$
\subfigure[$n-\alpha$ potentials.]{\includegraphics[width=7.5cm]{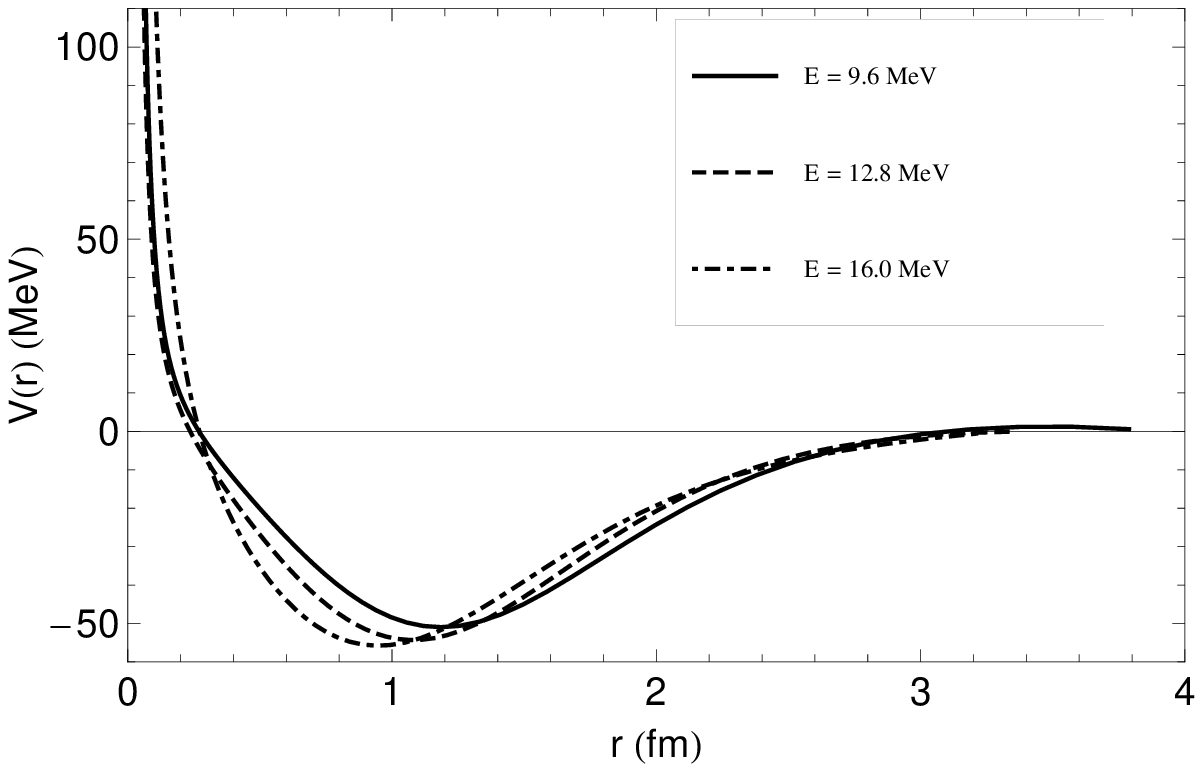}}
     \\
     \caption{Construction of potentials from experimental phase shifts listed in Table \ref{tabna}. (a) HA (continuous line) and mNS (dashed line) inverse potentials for $e-Ar$ scattering at 12 eV. (b) Inverse potentials for $n-\alpha$ scattering at three c.m. energies of $9.6$, $12.8$, and $16.0$ MeV. } \label{figna}
\end{center}
\end{figure}

The experimental phase shifts derived by Williams \cite{Willi} from $e-Ar$ scattering experiment at $E_{\rm c.m.}=12$ eV and the inversion parameters $a$ and $c$ are listed in the first line of Table \ref{tabna}. At $c=-3.7$ and $h=1.9$ the bound state parameter was found to be $\lambda=-0.44$. The resulted potential of the HA calculation is shown by the continuous line in Fig. \ref{figna}(a) and compared with that obtained by the modified Newton-Sabatier (mNS) inversion method \cite{Scheid} (dashed curve). Both potentials compare well showing the usability of the present method.
The $e-Ar$ potential has an attractive part with a minimum value of about $-2.8$ au at a distance of $r\approx 1.2$ au. At smaller distances the potential is of repulsive nature which can be interpreted as a manifestation of the Pauli-principle. Further details can be found in Ref. \cite{apagyi97}.

\begin{table}
\caption{\label{tabna} Input phase shifts $\delta_l$ derived from  $e-Ar$ and $n-\alpha$ scattering experiments performed at one and  three centre of mass energies, respectively. Corresponding inversion parameters $c$ and $a$.  $h=1.9$ was used in case of the $e-Ar$ inversion while $h=0$ was taken for the three  $n-\alpha$ inversions.}

\begin{center}
\begin{tabular}{rrrrrrr}
\hline
$E_{\rm c.m.}$ & $\delta_0$&$\delta_1$&$\delta_2$&$\delta_3$&$a$&$c$\\
\hline
$12$ (eV) &$-1.218$&$-0.626$&$1.191$&0.118&$3.9$&$-3.70$\\

$9.6$ (MeV) &$\phantom{-}1.763$&$\phantom{-}1.553$&$0.028$&&$3.9$&$-4.25$\\
$12.8$ (MeV) &$\phantom{-}1.676$&$\phantom{-}1.466$&$0.066$&&$3.4$&$-2.96$\\
$16.0$ (MeV) &$\phantom{-}1.588$&$\phantom{-}1.396$&$0.117$&&$3.3$&$-2.37$\\
\hline
\end{tabular}
\end{center}
\end{table}

\subsubsection{ $n-\alpha$ particle scattering.}
Input data at energies  $E_{\rm c.m.}=9.6, 12.8, $ and $16.0 $ MeV are listed  in the lower three lines of Table \ref{tabna} together with the inversion parameters $a$ and $c$.  The input phase shifts are taken from the comprehensive  analysis of the $n-\alpha$ scattering presented by Ali et al \cite{Ali}. Because of the spin-orbit coupling both spin-up $\delta_l^+$  and spin-down $\delta_l^-$ phase shifts contribute to the scattering amplitude at each partial wave. In case of weak spin-orbit coupling (which is assumed) the combined phase shifts
\begin{equation}
\delta_l=\frac{1}{2l+1}[(l+1)\delta_l^+ +l\delta_l^-]
\end{equation}
are characteristic of  the underlying central potential \cite{Leeb1995}, and these data are used as input for the HA procedure.

The resulting three HA potentials are exhibited in Fig. \ref{figna}(b). As we see they offer a similar physical interpretation for the scattering process as before in the case of electron-inert gas atom collision: the approaching colliding partners, neutron and alpha particle are attracting each other when entering the domain of nuclear forces. The attraction is culminating around the $\alpha-$particle surface between $r\approx 1.1-1.2$ fm, reaching a strength of potential energy between $-50$ and $-52$ MeV. When the colliding partners are merging the Pauli repulsion (originating from the fermionic exchange) takes into  effect and this is overcome by the nucleonic soft core repulsion at very small distances at $r\approx 0.1-0.2$ fm. Because the resulting inversion potentials are also very similar at these different energies between $9-16$ MeV, we may have found the energy-independent potential responsible for the scattering data which is always the desirable goal of any fixed energy inversion procedure.

\section{Conclusions}

We have surveyed, developed further and applied the constructive inverse scattering method of Horv\'ath and Apagyi (HA) \cite{HAMPLB}. The inverse problem consists of finding a scattering potential of finite support in the radial Schr\"odinger equation from a finite number of scattering  phase shifts given at one fixed energy. The solution procedure consists of solving an auxiliary inverse spectral problem of the  classical Sturm-Liouville equation whose spectral data is determined by the phase shifts of the fixed energy inverse scattering problem. The auxiliary inverse spectral problem is equivalent to a moment problem.

The HA method has been developed further in that two hidden parameters of the theory have been disclosed and enabled to vary. One of these parameters is a scale parameter $c$  appearing in the Liouville transformation. It plays an important role in the solution of the moment problem. The other parameter is an initial value parameter $h$ which enters the Gel'fand-Levitan constructive inversion scheme. The hidden (also called standard) values of these parameters have been $c=-1$ and $h=0$. By making these parameters  free the HA method proves to be applicable to cases where only a limited precision of input data is expected as in the case of the inversion of phase shifts derived from measured cross sections.

Depending on the number of bound states present in the auxiliary inverse spectral problem, various solution methods of the moment problem have been presented and applied to calculate the potential from a set of given phase shifts which  can be either calculated theoretically if the underlying potential is known (reconstruction procedure) or derived, e.g. from collision experiments if the potential is not known (construction procedure). Of course, an inverse method aims at solving the latter task, and the prior one may serve for testing or developing further the procedure.

 The examples illustrate that a dramatic improvement of the HA inversion
method can be achieved by the proper adjustment of the parameters $c$ and $h$. We may, for example,  reduce the number of bound states of the auxiliary inverse spectral problem by one. Because the most sensitive part of the method is the possible presence of auxiliary bound states we have established a procedure to assess their number. This procedure is based on the free motion because only the asymptotical part of the auxiliary potential is influenced by the wanted fixed energy potential. Alternatively, one may also use other (less sophisticated) fixed energy inverse scattering methods,  e.g. the modified Newton-Sabatier  procedure \cite{Scheid}, to assess  the number of auxiliary bound states. By using an independent phase equivalent method, one can simultaneously check the inverse potential provided by the HA procedure.

Finding an optimal choice of the parameters $a,c$, and $h$ can proceed through satisfying physical arguments. For example, one can obtain a cheap calculation (that is using small number of input data) via prescribing a maximal smoothness for the inverse potential. With such a prescription we have reproduced known potentials. Another (or simultaneous) prescription can be to check the inverse potential whether it fulfills the approximate relation $V(a)\approx 0$. With this prescription we have got the inverse potentials equivalent to the measured phase shifts of $e-Ar$ and $n-\alpha$ scattering experiments.
Note that by making the parameters $c$ and $h$ flexible one may also avoid the appearance of non-physical (e.g. singular) inverse potentials \cite{PAJMP} which can arise e.g. when the solution of the GL equation is not unique.

The HA method can be extended into various directions. A natural extension is to develop the theory to handle complex phase shifts which describe inelastic processes. Another extension would be to efficiently treat Coulombic processes, i.e. charged particle scattering. An interesting and important extension would be an alternative formulation where the auxiliary spectral problem is solved by the Marchenko integral equation \cite{marchenko}. The latter development is in progress.

\section*{Acknowledgements}
The authors thank Professor Mikl\'os Horv\'ath for valuable discussions and for reading the manuscript.

\end{document}